\documentclass[pre,twocolumn,showpacs,amsmath,amssymb]{revtex4}

\usepackage{graphicx}% Include figure files
\usepackage{dcolumn}% Align table columns on decimal point
\usepackage{bm}% bold math
\usepackage{cases}
\begin{document}

\preprint{APS/123-QED}

\title{Scale-free networks embedded in fractal space}% Force line breaks with \\

\author{K.~Yakubo}%
\email{yakubo@eng.hokudai.ac.jp}
\affiliation{Department of Applied Physics, Hokkaido University, Sapporo 060-8628, Japan}
\author{D.~Koro\v sak}
% \altaffiliation[]{}%Lines break automatically or can be forced with \\
\email{dean.korosak@uni-mb.si}
\affiliation{University of Maribor, Institute of Physiology, Faculty of Medicine,
Maribor SI-2000, Slovenia}
\affiliation{University of Maribor, Faculty of Civil Engineering, Maribor SI-2000, Slovenia}

\date{\today}% It is always \today, today,
             %  but any date may be explicitly specified

\begin{abstract}
The impact of inhomogeneous arrangement of nodes in space on
network organization cannot be neglected in most of real-world
scale-free networks. Here, we propose a model for a
geographical network with nodes embedded in a fractal space in
which we can tune the network heterogeneity by varying the
strength of the spatial embedding. When the nodes in such
networks have power-law distributed intrinsic weights, the
networks are scale-free with the degree distribution exponent
decreasing with increasing fractal dimension if the spatial
embedding is strong enough, while the weakly embedded networks
are still scale-free but the degree exponent is equal to
$\gamma=2$ regardless of the fractal dimension. We show that
this phenomenon is related to the transition from a non-compact
to compact phase of the network and that this transition accompanies
a drastic change of the network efficiency. We test our
analytically derived predictions on the real-world example of
networks describing the soil porous architecture.
\end{abstract}

\pacs{}% PACS, the Physics and Astronomy
                             % Classification Scheme.
%\keywords{Suggested keywords}%Use showkeys class option if keyword
                              %display desired
\maketitle

\section{Introduction}
\label{Introduction}

Scale-free organization of networks
\cite{barabasi2009,barabasi1999,albert2002} seems to be the
underlying principle common of many complex systems. In
real-world networks, examples include the Internet, social
networks or communication networks
\cite{yook2002,eubank2004,liben-nowell2005,gastner2006,lambiotte2008,bianconi2009},
the inhomogeneous arrangement of nodes in space strongly
impacts the network organization and the linking rules must
include the dependence on the distances between nodes
\cite{xulvi-brunet2002,barthelemy2003,kaiser2004,xulvi-brunet2007}.
However, most of the network models studied so far considered
either randomly distributed nodes in metric space
\cite{dall2002,herrmann2003,masuda2005,morita2006} or nodes
placed on a lattice
\cite{kleinberg2000,warren2002,ben-avraham2003,rozenfeld2002,kosmidis2008}.
The importance of inhomogeneous spatial positions of nodes was
emphasized in \cite{yook2002} where it was shown that the
fractality of the space is one of the universal parameters that
constrain the Internet models describing Internet's
large-scale topology and its observed scale-free character
\cite{faloutsos1999,vazquez2002,siganos2003} at the router and
autonomous system level. While preferential attachment
\cite{barabasi1999,albert2002} seems to be the main underlying
mechanism structuring the Internet, the original form of the
preferential attachment \cite{barabasi1999} should be altered
\cite{yook2002,vazquez2002,shakkottai2010} to account for the
observed spatial and/or functional heterogeneity of the nodes.

Here, we study the structure of networks formed by geographical
network model in which the nodes with power-law distributed
intrinsic weights (i.e., fitnesses \cite{caldarelli2002,masuda2004,garlaschelli2007})
are embedded in a fractal space. We show analytically that the
networks produced by such model are scale-free with the degree
exponent influenced by the fractal dimension of the embedding
space if the spatial embedding is strong enough. By explicitly
deriving the degree and the edge-length distribution functions,
we classify these networks into non-compact phase with
infinite average degree and average edge length and compact
phase with finite average degree and average edge length,
separated by an intermediate phase characterized by finite
average degree and infinite average edge length. It is also
shown that the transition between these phases accompanies a
drastic change of the network efficiency. Finally, we use our
findings in the analysis of the networks describing the soil
porous architecture as an example.

\section{Model}
\label{Model}

In order to describe an inhomogeneous distribution of $N$ nodes
in a space of linear dimension $L$, the nodes are isotropically
distributed in a fractal manner with the fractal dimension $D$.
In the theoretical treatment we employ an \textit{artificial}
boundary conditions for which every node position is
statistically equivalent. This is slightly different from usual
periodic boundary conditions leading to an anisotropy. The
number of nodes is thus given by $N=\rho\Omega\int_{0}^{L}
l^{D-1}\,dl$, where $\Omega$ is the $D$-dimensional solid angle
and $\rho$ is the density of nodes. Each node has a real and
continuous fitness $x$ randomly assigned according to the
distribution function $s(x)$. If a node pair $(i$-$j)$
satisfies
\begin{equation}
\frac{F(x_{i},x_{j})}{l_{ij}^{m}}>\Theta\ ,
\label{condition1}
\end{equation}
then these two nodes are connected by an edge, where $l_{ij}$
is the Euclidean distance between the nodes $i$ and $j$, $m
(\ge 0)$ is a real parameter quantifying the strength of the
spatial embedding (namely, the strength of the geographical
effect), $\Theta$ is a threshold value, and $F(x,y)$ is a
function relating the two fitnesses to the connectivity. If
$m=0$, we have a conventional fitness model which provides a
scale-free network for a variety of combinations of forms of
$F(x,y)$ and $s(x)$ \cite{caldarelli2002,masuda2004}. Here, we
concentrate on the case of
\begin{equation}
F(x,y)=xy\ ,
\label{Fxy}
\end{equation}
and
\begin{equation}
s(x)=s_{0}x^{-\alpha}\ ,
\label{sx}
\end{equation}
where $x$ (and $y$) are in the range of $[x_{\text{min}},\infty)$
and $\alpha>1$. From the normalization condition of $s(x)$, we
have
\begin{equation}
s_{0}=(\alpha-1)x_{\text{min}}^{\alpha-1}\ .
\label{s0}
\end{equation}
Similar geographical network models have been studied under the
assumption that nodes are homogeneously distributed in a
Euclidean space \cite{masuda2005,morita2006}, while here we
investigate how the inhomogeneity (fractality) of spatial
distribution of nodes affects properties of the network.

\section{Degree distribution function}
\label{Degree}

First, we calculate the degree distribution function of the
network formed by the above geographical algorithm. Let
$k_{i}(l)dl$ be the number of nodes connected to the node $i$
and included in a thin spherical shell of the radius $l$ (width
$dl$) centered at the position of the node $i$. Since the
distance between the node $i$ and a node in the shell is $l$,
the connectivity condition Eq.~(\ref{condition1}) tells us that
nodes with $x_{j}>\Theta l^{m}/x_{i}$ can connect to the node
$i$. Thus, the number of connected nodes $k_{i}(l)dl$ is
\begin{equation}
k_{i}(l)dl=n(l)dl\cdot \int_{\Theta l^{m}/x_{i}}^{\infty}s(x)\,dx\ ,
\label{kil1}
\end{equation}
where $n(l)dl=\rho\Omega l^{D-1}dl$ is the number of nodes in
this spherical shell. In this expression, we assume that
$\Theta l^{m}/x_{j}>x_{\text{min}}$, which is equivalent to
$l>l_{\text{min}}(x_{i})$ where
\begin{equation}
l_{\text{min}}(x_{i})=\left(\frac{x_{\text{min}}x_{i}}{\Theta}\right)^{1/m}\ .
\label{lmin}
\end{equation}
Nodes within the distance $l_{\text{min}}(x_{i})$ from the
node $i$ can connect to the node $i$. Thus, using Eq.~(\ref{s0}),
we have
\begin{widetext}
\begin{subnumcases}
{k_{i}(l)= \label{kil2}}
\displaystyle
\rho\Omega\left(\frac{x_{\text{min}}}{\Theta}\right)^{\alpha-1}x_{i}^{\alpha-1}
l^{D-1-m(\alpha-1)} & ,\ $l>l_{\text{min}}(x_{i})$ , \label{kil2a}\\[2mm]
\rho\Omega l^{D-1}  &,\  $l\le l_{\text{min}}(x_{i})$\ .
\label{kil2b}
\end{subnumcases}
\end{widetext}
Integrating $k_{i}(l)$ with respect to $l$ over $(0,L]$ we
obtain the total number of nodes connected to the node $i$, or
the degree $k_{i}$ of the node $i$ given by
\begin{eqnarray}
k_{i}&=& \frac{\rho\Omega x_{\text{min}}^{\alpha-1}L^{D-m(\alpha-1)}}
{[D-m(\alpha-1)]\Theta^{\alpha-1}} x_{i}^{\alpha-1} \nonumber \\
&&+\frac{\rho\Omega x_{\text{min}}^{D/m}}{\Theta^{D/m}}
\left[\frac{1}{D}-\frac{1}{D-m(\alpha-1)}\right] x_{i}^{D/m}.
\label{ki}
\end{eqnarray}
It should be noted that geometrical coefficients in two terms
of Eq.~(\ref{ki}) coming from the volume integration are not
very meaningful for realistic systems because of our artificial
boundary conditions. In the case of $D-m(\alpha-1)>0$, the
first term of Eq.~(\ref{ki}) dominates the second term when $L$
is sufficiently large and we have $k_{i}\propto
x_{i}^{\alpha-1}$. The degree distribution function $P(k)$
calculated from the relation $|P(k)dk|=|s(x)dx|$ is then
proportional to $k^{-2}$ independently of $m$, $\alpha$, and
$D$. If however, $D-m(\alpha-1)\le 0$, the second term of
Eq.~(\ref{ki}) becomes much larger than the first term. The
relation $k_{i}\propto x_{i}^{D/m}$ in this case leads to
$P(k)\propto k^{-m(\alpha-1)/D+1}$. Therefore, the degree
distribution function of the present geographical network model
is given by
\begin{subnumcases}
{P(k)\propto \label{Pk1}}
k^{-2}                       & ,\ $m\le m_{\text{c}0}$ \label{Pk1a} ,\\
\displaystyle
k^{-m(\alpha-1)/D+1}        & ,\ $m >  m_{\text{c}0}$ \ ,
\label{Pk1b}
\end{subnumcases}
where
\begin{equation}
m_{\text{c}0}=\frac{D}{\alpha-1}\ .
\label{mc0}
\end{equation}
The distribution function $P(k)$ obeys power-law forms in both
cases of $m\le m_{\text{c}0}$ and $m> m_{\text{c}0}$, and the
degree exponent $\gamma$ is
\begin{equation}
\gamma=
\begin{cases}
2                       & ,\ m\le m_{\text{c}0} \ ,\\
\displaystyle
\frac{m}{D}(\alpha-1)+1 & ,\ m >  m_{\text{c}0} \ .
\end{cases}
\label{gamma}
\end{equation}
The geographical inhomogeneity of the node distribution does
not affect the degree distribution of the network for a weak
geographical effect (small $m$), while $\gamma$ depends on $D$
for a strong effect (large $m$).

In the above argument, we assumed that the system size $L$ is
always larger than $l_{\text{min}}(x_{i})$ for any $x_{i}$,
because we are interested in the thermodynamic case
($L\to\infty$). This is, however, not obvious, because $x_{i}$
[and then $l_{\text{min}}(x_{i})$] can also diverge in the
thermodynamic limit under a constant density $\rho$. Let us
consider carefully this condition $L>l_{\text{min}}(x_{i})$. In
a finite system with $N=\rho\Omega L^{D}/D$ nodes, the fitness
$x$ is truncated at a finite value. The maximum fitness
$x_{\text{max}}$ is given by
$N\int_{x_{\text{max}}}^{\infty}s(x)\,dx=1$. Using the fitness
distribution Eq.~(\ref{sx}) with Eq.~(\ref{s0}), the quantity
$x_{\text{max}}$ is given by
\begin{equation}
x_{\text{max}}=N^{1/(\alpha-1)}x_{\text{min}}\ .
\label{xmax}
\end{equation}
Thus, the length $l_{\text{min}}(x_{i})$ can be as large as
$l_{\text{min}}(x_{\text{max}})$, where
\begin{equation}
l_{\text{min}}(x_{\text{max}}) =
\left[\frac{(\rho\Omega)^{1/(\alpha-1)}L^{D/(\alpha-1)}x_{\text{min}}^{2}}{\Theta D^{1/(\alpha-1)}}\right]^{1/m}.
\label{lmaxxmax}
\end{equation}
From the above expression, the condition $l_{\text{min}}(x_{\text{max}})<L$
is equivalent to $\Theta>\Theta_{0}$, where
\begin{equation}
\Theta_{0}=\left(\frac{\rho\Omega}{D}\right)^{\frac{1}{\alpha-1}}x_{\text{min}}^{2}L^{\frac{D}{\alpha-1}-m}\ .
\label{theta0}
\end{equation}
If $m>m_{\text{c}0}$, the quantity $\Theta_{0}$ goes to zero in
the thermodynamic limit and any finite $\Theta$ satisfies
$\Theta>\Theta_{0}$, namely,
$l_{\text{min}}(x_{\text{max}})<L$. Thus, the degree
distribution function $P(k)$ is given by Eq.~(\ref{Pk1b}) in
the thermodynamic limit with $m>m_{\text{c}0}$. On the
contrary, if $m\le m_{\text{c}0}$, $\Theta$ is always less than
$\Theta_{0}$ because $\Theta_{0}$ diverges as $L\to\infty$. In
this case, there must be nodes satisfying
$l_{\text{min}}(x_{i})>L$ for which $x_{i}>\Theta
L^{m}/x_{\text{min}}$. Since the condition $x_{i}>\Theta
L^{m}/x_{\text{min}}$ implies that any node in the whole system
can connect to the node $i$, the degree of such a node is $N-1$
independent of $x_{i}$. Thus, these nodes give an additional
$\delta$-functional contribution [$\delta(k-N+1)$] to the
degree distribution $P(k)$ given by Eq.~(\ref{Pk1a}) for nodes
with $x_{i}<\Theta L^{m}/x_{\text{min}}$. Let us estimate the
magnitude of this $\delta$-functional contribution. It is
proportional to the number of nodes $n_{0}$ having
$x_{i}>\Theta L^{m}/x_{\text{min}}$. The quantity $n_{0}$ is
given by $N\int_{\Theta L^{m}/x_{\text{min}}}^{\infty}
s(x)\,dx$ and can be written as
\begin{equation}
n_{0}= N\left(\frac{L}{\xi}\right)^{-m(\alpha-1)} \ ,
\label{n0}
\end{equation}
where $\xi$ is the node-pair distance defined by
\begin{equation}
\xi=l_{\text{min}}(x_{\text{min}})=\left(\frac{x_{\text{min}}^{2}}{\Theta}\right)^{1/m} \ ,
\label{xi}
\end{equation}
below which the two nodes are connected independently of the fitness.
The properly normalized $\delta$-functional part of $P(k)$ is
then presented by $(L/\xi)^{-m(\alpha-1)}\delta(k-N+1)$. Since
$m(\alpha-1)$ is always positive, the $\delta$-functional
contribution vanishes in the thermodynamic limit. Therefore,
Eq.~(\ref{Pk1}) is valid both for the cases of $m\le
m_{\text{c}0}$ and $m> m_{\text{c}0}$ in an infinite system.
For a finite $L$, however, $\Theta$ can be chosen to be less
than $\Theta_{0}$ independently of $m$ and the
$\delta$-functional contribution remains finite. Thus, the
degree distribution function of a finite system with
$\Theta<\Theta_{0}$ must have the $\delta$-functional
correction term, that is,
\begin{equation}
P(k)=p_{0}k^{-\gamma}+\left(\frac{L}{\xi}\right)^{-m(\alpha-1)}\delta(k-N+1)\ ,
\label{Pk2}
\end{equation}
for both $m\le m_{\text{c}0}$ and $m> m_{\text{c}0}$. Here,
$p_{0}$ is a normalization constant and the exponent $\gamma$
is given by Eq.~(\ref{gamma}). It should be noted that
Eq.~(\ref{Pk2}) holds for a finite but large $L$ because the
first term of Eq.~(\ref{Pk2}) is valid in the large $L$ limit.

The quantity $\Theta_{0}$ is a characteristic value of $\Theta$
peculiar to a finite system with a fixed density $\rho$. There
are two other characteristic values of $\Theta$ for a finite
$L$. One is $\Theta_{\text{min}}$ below which every node can
connect to all other $N-1$ nodes. Network constructed under
$\Theta<\Theta_{\text{min}}$ becomes the complete graph.
Obviously, $\Theta_{\text{min}}$ is given by
\begin{equation}
\Theta_{\text{min}}=\frac{x_{\text{min}}^{2}}{L^{m}}\ .
\label{thetamin}
\end{equation}
Another characteristic $\Theta$ is $\Theta_{\text{max}}$
above which no node can connect to any other nodes. Network
with $\Theta>\Theta_{\text{max}}$ is a set of isolated nodes.
We have
%The quantity $\Theta_{\text{max}}$ is expressed as
%
\begin{equation}
\Theta_{\text{max}}=\frac{x_{\text{max}}^{2}}{\Delta l^{m}}\ ,
\label{thetamax1}
\end{equation}
where $\Delta l=(\rho \Omega/D)^{-1/D}$ is the minimum edge
length. Using Eq.~(\ref{xmax}), $\Theta_{\text{max}}$ is
written as
\begin{equation}
\Theta_{\text{max}}=\left(\frac{\rho\Omega}{D}\right)^{\frac{2}{\alpha-1}+\frac{m}{D}}
x_{\text{min}}^{2}L^{\frac{2D}{\alpha-1}}\ .
\label{thetamax2}
\end{equation}
In a finite system we always assume that $\Theta$ satisfies the
condition $\Theta_{\text{min}}\ll \Theta\ll\Theta_{\text{max}}$
leading to non-trivial networks. It is not necessary to
consider this condition in an infinite system because
$\Theta_{\text{min}}$ and $\Theta_{\text{max}}$ vanishes and
diverges, respectively, in the thermodynamic limit.

\section{Relation between $\langle k\rangle$ and $\Theta$}
\label{Relation}

From Eq.~(\ref{Pk1}), it is clear that the average degree
$\langle k\rangle$ diverges for $m\le m_{\text{c}0}$ because of
$\gamma=2$ and it remains finite for $m> m_{\text{c}0}$ in the
thermodynamic limit. The average degree $\langle k\rangle$ in a
finite system is, however, always finite and depends on
$\Theta$. A large $\Theta$ restricts a connection of a node
pair and leads a small $\langle k\rangle$. It is important to
know the relation between $\Theta$ and $\langle k\rangle$ for a
finite but large system. We calculate the average degree by
\begin{equation}
\langle k\rangle=\int_{x_{\text{min}}}^{x_{\text{max}}} k_{i}s(x)\,dx\ ,
\label{kav1}
\end{equation}
instead of $\langle k\rangle=\int kP(k)dk$, using the derived
expression (not asymptotic form) of $k_{i}$ for a finite
system.

Let us consider Eq.~(\ref{kav1}) separately for
$\Theta<\Theta_{0}$ and for $\Theta\ge \Theta_{0}$. In the case of
$\Theta<\Theta_{0}$, the length $l_{\text{min}}(x_{i})$ can be larger
than $L$, which implies $\Theta L^{m}/x_{\text{min}}<x_{\text{max}}$.
Thus, the integral of Eq.~(\ref{kav1}) is separated into two regions
\begin{equation}
\langle k\rangle=\int_{x_{\text{min}}}^{\Theta L^{m}/x_{\text{min}}} k(x)s(x)\,dx
+(N-1)\int_{\Theta L^{m}/x_{\text{min}}}^{x_{\text{max}}}s(x)\,dx\ ,
\label{kav2}
\end{equation}
where $k(x)$ is given by Eq.~(\ref{ki}) regarding $x_{i}$ as a
continuous variable $x$. The coefficient $N-1$ in the second term
comes from the fact that nodes satisfying $x_{i}>\Theta L^{m}/x_{\text{min}}$
connect to all $N-1$ nodes. Using Eq.~(\ref{ki}) and approximating
$\int_{\Theta L^{m}/x_{\text{min}}}^{x_{\text{max}}}dx$ by
$\int_{\Theta L^{m}/x_{\text{min}}}^{\infty}dx$, we have
\begin{eqnarray}
\langle k\rangle&=&X\rho\left(\frac{x_{\text{min}}^{2}}{\Theta}\right)^{\alpha-1}L^{D-m(\alpha-1)}\log
\left(\frac{Y\Theta L^{m}}{x_{\text{min}^{2}}}\right)\nonumber \\
&&+Z\rho\left(\frac{x_{\text{min}}^{2}}{\Theta}\right)^{D/m}\ ,
\label{kav3}
\end{eqnarray}
where
\begin{eqnarray}
X&=&\frac{\Omega(\alpha-1)}{D-m(\alpha-1)}\ ,
\label{X}
\\
Y&=&\exp\left\{\frac{d-2m(\alpha-1)}{(\alpha-1)[d-m(\alpha-1)]}\right\}\ ,
\label{Y}
\\
Z&=&\frac{\Omega m^{2}(\alpha-1)^{2}}{D[D-m(\alpha-1)]^{2}} \ .
\label{Z}
\end{eqnarray}
It should be again emphasized that these geometrical quantities
$X$, $Y$, and $Z$ resulting from the volume integration depend
strongly on the boundary conditions and are not very
meaningful. We can evaluate the asymptotic behavior of $\langle
k\rangle$ for large $L$ by using Eq.~(\ref{kav3}). In the case
of $m<m_{\text{c}0}$, the first term of Eq.~(\ref{kav3})
obviously dominates the second term. Then, ignoring unimportant
geographical coefficient, $\langle k\rangle$ behaves
asymptotically as
\begin{equation}
\langle k\rangle \sim \Theta^{1-\alpha}(\log \Theta+c)\ .
\label{kav4}
\end{equation}
where $c$ is a constant depending on the boundary condition. On
the other hand, a careful treatment is required for
$m>m_{\text{c}0}$. It seems that the second term of
Eq.~(\ref{kav3}) dominates the first term for $m>m_{\text{c}0}$
in the thermodynamic limit. However, we should note that
$\Theta$ must be infinitesimal to satisfy the condition
$\Theta<\Theta_{0}$ in this calculation because $\Theta_{0}$
for $L\to\infty$ goes to zero for $m>m_{\text{c}0}$ so it is
not obvious which term is dominating in Eq.~(\ref{kav3}). In
order to find the dominant term, we evaluate the lower bounds
of these terms by replacing $\Theta$ by $\Theta_{0}$ and
using~(\ref{theta0}). Then, the lower bounds of the first and
the second terms are proportional to $\log [Y(\rho
\Omega/D)^{1/(\alpha-1)}L^{D/(\alpha-1)}]$ and
$L^{\left[1-\frac{D}{m(\alpha-1)}\right]}$ respectively. This
suggests that the second term dominates the first term for
large $L$ because the exponent $1-\frac{D}{m(\alpha-1)}$ is
positive for $m>m_{\text{c}0}$. We have then
\begin{equation}
\langle k\rangle \sim \Theta^{-D/m}\ ,
\label{kav5}
\end{equation}
for $\Theta<\Theta_{0}$ and $m>m_{\text{c}0}$. At $m=m_{\text{c}0}$, both
terms in Eq.~(\ref{kav3}) should be considered.

Next, we treat the case of $\Theta\ge \Theta_{0}$. Here
$l_{\text{min}}(x_{i})$ is always less than $L$, and the degree
$k_{i}$ is given by Eq.~(\ref{ki}) for any $x_{i}$ in the
range of $x_{\text{min}}\le x_{i}\le x_{\text{max}}$. The average
degree $\langle k\rangle$ is then simply presented by
\begin{eqnarray}
\langle k\rangle &=&
\int_{x_{\text{min}}}^{x_{\text{max}}} k(x)s(x)\,dx \nonumber \\
&=& \frac{\rho X}{\alpha-1}\left(\frac{x_{\text{min}}^{2}}{\Theta}\right)^{\alpha-1}
L^{D-m(\alpha-1)} \log \left(W L^{D}\right) \nonumber \\
&& -\rho Z[D-m(\alpha-1)]\left(\frac{x_{\text{min}}^{2}}{\Theta}\right)^{D/m}
\left(W L^{D}\right)^{\frac{D}{m(\alpha-1)}-1} \nonumber \\
&& +\rho Z[D-m(\alpha-1)]\left(\frac{x_{\text{min}}^{2}}{\Theta}\right)^{D/m}\ ,
\label{kav6}
\end{eqnarray}
where $W=\rho\Omega/D$ and we used Eq.~(\ref{ki}) for $k(x)$
and Eq.~(\ref{xmax}). For $m>m_{\text{c}0}$, it is easy to
understand that the third term dominates other two terms for
large $L$. So we have
\begin{equation}
\langle k\rangle \sim \Theta^{-D/m} \ .
\label{kav7}
\end{equation}
In the case of $m<m_{\text{c}0}$, the infinitely large $\Theta$
must be considered when we find the dominant term of
Eq.~(\ref{kav6}), because $\Theta$ is larger than $\Theta_{0}$
and $\Theta_{0}$ diverges for $m<m_{\text{c}0}$ in the
thermodynamic limit. As in the case of $\Theta<\Theta_{0}$ and
$m>m_{\text{c}0}$, replacing $\Theta$ in Eq.~(\ref{kav6}) by
$\Theta_{0}$, the $L$ dependence of the upper bounds of these
terms points to the dominant term. Since the upper bounds of
the first, second, and third terms are proportional to
$\log(\rho \Omega L^{D}/D)$, $L^{0}$, and
$L^{D[1-\frac{1}{m(\alpha-1)}]}$, respectively, the first term
dominates the second and third terms because of
$m<m_{\text{c}0}$. Therefore, the average degree $\langle
k\rangle$ is asymptotically given by
\begin{equation}
\langle k\rangle \sim \Theta^{1-\alpha} \ .
\label{kav8}
\end{equation}
This relation differs from Eq.~(\ref{kav4}) by a logarithmic
correction.

In summary, the relation between $\Theta$ and $\langle k\rangle$
is given by
\begin{subnumcases}
{\langle k\rangle \sim \label{kavresult1}} \Theta^{1-\alpha}(\log \Theta +c) & , $m<m_{c0}$
\label{kavresult11}
\\
\Theta^{-D/m} & , $m>m_{c0}$
\label{kavresult12}
\end{subnumcases}
for $\Theta<\Theta_{0}$ and
\begin{subnumcases}
{\langle k\rangle \sim \label{kavresult2}}
\Theta^{1-\alpha} & , $m<m_{c0}$
\label{kavresult21}
\\
\Theta^{-D/m} & , $m>m_{c0}$\ .
\label{kavresult22}
\end{subnumcases}
for $\Theta\ge \Theta_{0}$. At $m=m_{\text{c}0}$, $\langle
k\rangle$ is related to $\Theta$ through Eq.~(\ref{kav3}) for
$\Theta<\Theta_{0}$ and through Eq.~(\ref{kav6}) for $\Theta\ge
\Theta_{0}$, because every term contributes equally to $\langle
k\rangle$ even in the thermodynamic limit. A similar result to
Eq.~(\ref{kavresult2}) has been obtained by \cite{morita2006}
where the nodes were uniformly distributed in a $d$-dimensional
space with the $L$-max norm and $m$ is fixed at $m=D(=d)$. The
asymptotic $L$ dependence of $\langle k\rangle$ for a fixed
$\Theta$ can be also evaluated from Eqs.~(\ref{kav3}) and
(\ref{kav6}). In the case of $m<m_{\text{c}0}$ and enough large
$L$ (then, $\Theta<\Theta_{0}$), the dominant first term of
Eq.~(\ref{kav3}) gives $\langle k\rangle \propto
L^{D-m(\alpha-1)}\log L$. For $m>m_{\text{c}0}$ (and then
$\Theta>\Theta_{0}$), we have $\langle k\rangle \propto L^{0}$.
These $L$ dependences are consistent with those calculated by
$\langle k\rangle=\int_{1}^{N-1} kP(k)\,dk$ by using
Eqs.~(\ref{Pk2}) and (\ref{Pk1b}) for $m<m_{\text{c}0}$ and
$m>m_{\text{c}0}$, respectively, and taking into account the
$L$ dependence of $p_{0}$ in Eq.~(\ref{Pk2}).

\section{Edge-length distribution function}
\label{Edge-length}

In this section, we derive the edge-length distribution
function $R(l)$ of our geographical networks embedded in a
fractal space. To this end, we regard $k_{i}(l)$ given by
Eq.~(\ref{kil2}) as a continuous function $k(x,l)$ of the
fitness $x$ and the edge length $l$. The average number of
edges, $k(l)dl$, of length $[l,l+dl]$ from a given node is
obtained by averaging $k(x,l)$ over the fitness $x$, i.e.,
\begin{equation}
k(l)=\int_{x_{\text{min}}}^{x_{\text{max}}}s(x)k(x,l)\,dx.
\label{kl1}
\end{equation}
Equation (\ref{kil2}) expresses the forms of $k_{i}(l)$ by
separating two cases $l>l_{\text{min}}(x_{i})$ and $l\le
l_{\text{min}}(x_{i})$ for a fixed $x_{i}$. Corresponding to
this classification, $k(x,l)$ in Eq.~(\ref{kl1}) for a fixed
$l$ has different forms for $x_{\text{min}}\le x<\Theta
l^{m}/x_{\text{min}}$ and $\Theta l^{m}/x_{\text{min}}\le x\le
x_{\text{max}}$, respectively. Thus, the integral of
Eq.~(\ref{kl1}) is calculated as
\begin{eqnarray}
k(l)&=& \rho\Omega\left(\frac{x_{\text{min}}}{\Theta}\right)^{\alpha-1}
l^{D-1-m(\alpha-1)}\int_{x_{\text{min}}}^{x_{l}}
s(x)x^{\alpha-1}\,dx \nonumber \\
&&+\rho \Omega l^{D-1}\int_{x_{l}}^{\infty}
s(x)\,dx\ ,
\label{kl2}
\end{eqnarray}
where $x_{l}=\Theta l^{m}/x_{\text{min}}$. Here, $l$ is assumed
to be larger than $\xi$ (namely, $x_{\text{min}}<x_{l}$) and
the upper cut-off of the integral is, as an approximation,
extended to infinity. Using Eqs.~(\ref{sx}), (\ref{s0}),
(\ref{kil2}), and (\ref{xi}), $k(l)$ is expressed by
\begin{equation}
k(l)=\rho \Omega l^{D-1}\left(\frac{l}{\xi}\right)^{-m(\alpha-1)}
\left[1+m(\alpha-1)\log\left(\frac{l}{\xi}\right) \right]\ .
\label{kl3}
\end{equation}
The probability distribution function $R(l)$ is given by
\begin{equation}
R(l)=\frac{k(l)}{\displaystyle \int_{L}k(l')dl'}\ ,
\label{rl1}
\end{equation}
where the integration in the denominator is done over the
whole range of $l$. Neglecting the normalization constant,
the edge-length distribution is
\begin{equation}
R(l)\propto l^{D-1}\left(\frac{l}{\xi}\right)^{-m(\alpha-1)}
\left[1+m(\alpha-1)\log\left(\frac{l}{\xi}\right) \right]\ .
\label{rl2}
\end{equation}
We should remark that $R(l)$ for $m\le m_{\text{c}0}$ goes to
infinity as $l\to \infty$. Since the length $l$ does not exceed
the size $L$ in a finite system, the distribution $R(l)$ is
actually truncated at $l=L$. In the thermodynamic limit, however,
we must consider the infinitesimal normalization constant coming
from the denominator of Eq.~(\ref{rl1}).

In the case of $l\le \xi$ (namely, $x_{\text{min}}\ge x_{l}$),
$k(x,l)$ in Eq.~(\ref{kl1}) is given by Eq.~(\ref{kil2b}) for
any $x$ in the integration range $[x_{\text{min}},x_{\text{max}}]$.
Thus, $k(l)$ is given by $\rho\Omega l^{D-1}\int_{x_{\text{min}}}^{\infty}s(x)dx$,
namely,
\begin{equation}
k(l)=\rho\Omega l^{D-1}\ ,
\label{kl4}
\end{equation}
and then,
\begin{equation}
R(l)\propto l^{D-1}\ .
\label{rl3}
\end{equation}
It should be noted that the distribution $R(l)$ depends on the fractal
dimension $D$ independently of $m$ while the degree distribution $P(k)$
does not depend on $D$ for $m\le m_{\text{c}0}$ (weak geographical effect
region).

Two expressions Eqs.~(\ref{rl2}) and (\ref{rl3}) of $R(l)$ for
$l>\xi$ and $l\le \xi$ must coincide at $l=\xi$. This condition concludes
that the proportionality coefficients for Eqs.~(\ref{rl2}) and (\ref{rl3})
are identical. We can derive the common coefficient $C$ for an infinite
system from the normalization condition of $R(l)$ given by
\begin{eqnarray}
1&=& C\int_{0}^{\xi} l^{D-1}dl \nonumber \\
&&+C\int_{\xi}^{\infty} l^{D-1}\left(\frac{l}{\xi}\right)^{-\beta}
\left[1+\beta\log\left(\frac{l}{\xi}\right)\right]dl\ ,
\label{normalization}
\end{eqnarray}
where $\beta=m(\alpha-1)$. When $m>m_{\text{c}0}$, this equation provides
a finite coefficient expressed by
\begin{equation}
C=\frac{D}{\xi^{D}}\left(1-\frac{m_{\text{c}0}}{m}\right)^{2}\ ,
\label{C}
\end{equation}
while $C$ is infinitesimal for $m\le m_{\text{c}0}$ as
mentioned above. It should be noted that the coefficient $C$
does not depend on the boundary condition because the
boundary-condition dependent factors in the numerator and the
denominator in Eq.~(\ref{rl1}) are canceled out.

From the above argument, we can immediately derive the
probability $g(l)$ of two nodes with distance $l$ to be
connected. This probability is given by the ratio of the number
of connected nodes $k(l)dl$ to $n(l)dl$ nodes located at
distances $[l,l+dl]$ from a given node, namely,
$g(l)=k(l)/n(l)$. Since $n(l)=\rho\Omega l^{D-1}$ as mentioned
below Eq.~(\ref{kil1}) and $k(l)$ is given by Eq.~(\ref{kl3})
or (\ref{kl4}) for $l>\xi$ or $l\le \xi$ respectively, we have
\begin{equation}
g(l)=
\begin{cases}
\displaystyle
\left(\frac{l}{\xi}\right)^{-m(\alpha-1)} \left[1+m(\alpha-1)\log\left(\frac{l}{\xi}\right) \right]
& ,\ l>\xi \ ,\\[4mm]
1 & ,\ l\le \xi \ .
\end{cases}
\label{gl1}
\end{equation}
This expression can be alternatively derived directly from
the meaning of $g(l)$,
\begin{equation}
g(l)=\int_{x_{\text{min}}}^{\infty}s(x)dx\int_{xy/l^{m}>\Theta}s(y)dy\ .
\label{gl2}
\end{equation}
Considering that two nodes are always connected if the fitness
of one node exceeds $x_{l}(=\Theta l^{m}/x_{\text{min}})$,
we can separate the above integration into two parts as
\begin{equation}
g(l)= \int_{x_{l}}^{\infty} s(x)dx + \int_{x_{\text{min}}}^{x_{l}} s(x)dx
\int_{\Theta l^{m}/x}^{\infty} s(y)dy \ ,
\label{gl3}
\end{equation}
for $x_{l}>x_{\text{min}}$ ($l>\xi$). For $x_{l}\le
x_{\text{min}}$, the integral range of the second integral in
Eq.~(\ref{gl2}) is spread over the whole region of $y$, then
$g(l)=\int_{x_{\text{min}}}^{\infty}
\int_{x_{\text{min}}}^{\infty} s(x)s(y)dxdy$. These equations
again lead Eq.~(\ref{gl1}) if we use Eq.~(\ref{sx}). We should
note that the probability $g(l)$ given by Eq.~(\ref{gl1}) does
not depends on the fractal dimension $D$ for any value of $m$.

\section{Numerical confirmations}
\label{Numerical}

\begin{figure}[ttt]
\includegraphics[width=0.38\textwidth]{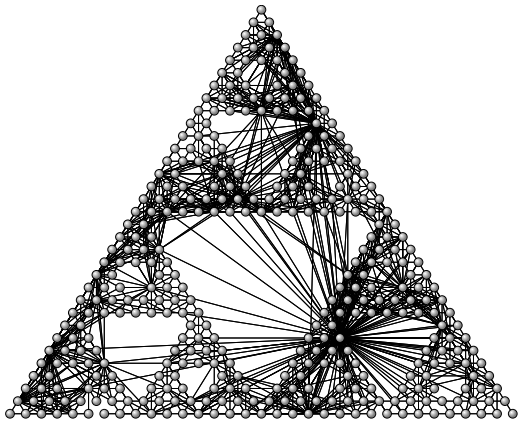}
\caption{\label{fig1}
Typical Sierpinski geographical network. Nodes are located on vertices of
the Sierpinski gasket in the 6th generation ($N=366$). Parameters to form
the network are $\alpha=2.0$ and $m=3.0$. The threshold $\Theta$ is
chosen to satisfy $\langle k\rangle=10.0$.}
\end{figure}
\begin{figure}[ttt]
\includegraphics[width=0.45\textwidth]{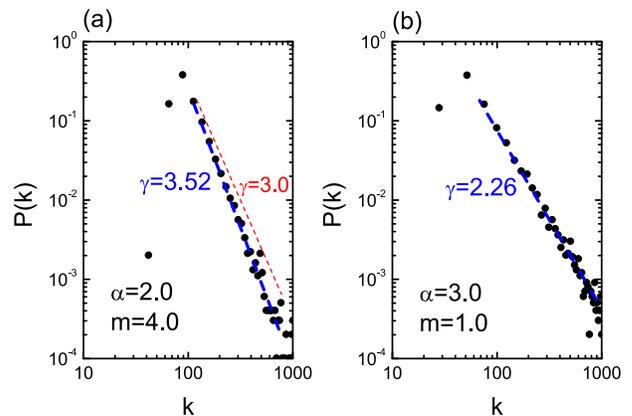}
\caption{\label{fig2}
(Color online) Degree distribution of Sierpinski geographical
networks in the 9th generation ($N=9,843$). The threshold
$\Theta$ is chosen to satisfy $\langle k\rangle=120.0$ for good
statistics. Parameters $\alpha$ and $m$ are set as (a)
$\alpha=2.0$ and $m=4.0$ and (b) $\alpha=3.0$ and $m=1.0$.
Thick dashed lines through dots indicate the slopes $\gamma$
predicted by Eq.~(\ref{gamma}). Thin dashed line in Fig.~2(a)
represents the slope calculated by using the Euclidean
dimension $d=2$ instead of the fractal dimension $D=1.585$.}
\end{figure}
\begin{figure}[bbb]
\includegraphics[width=0.45\textwidth]{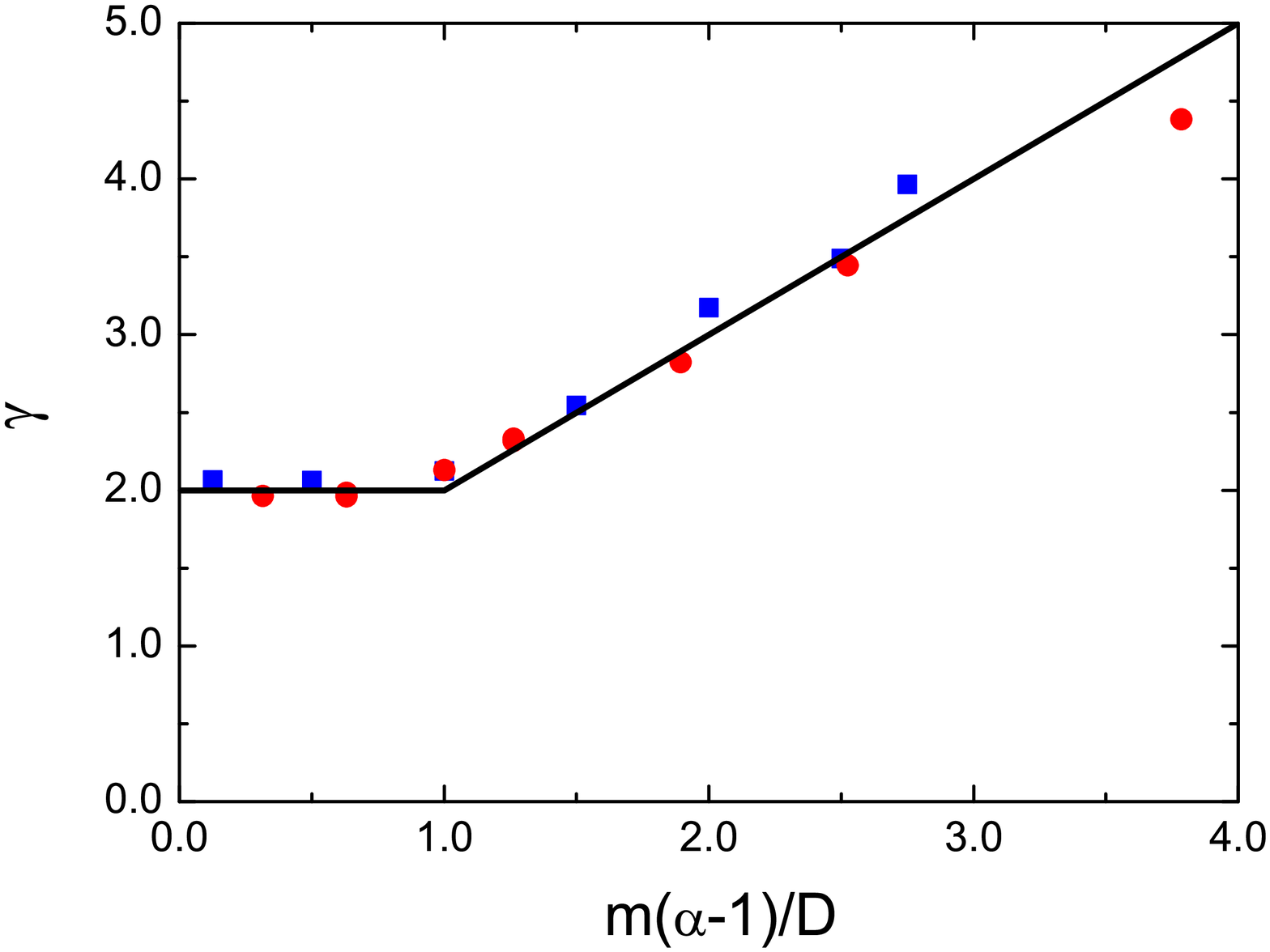}
\caption{\label{fig3}
(Color online) Degree exponent $\gamma$ as a function of
$m(\alpha-1)/D$. Solid line shows the theoretical prediction
Eq.~(\ref{gamma}). Filled circles and squares represent
numerically obtained $\gamma$ for Sierpinski geographical
networks (9th generation) and 2d-geographical networks
($N=1,000$) with several combinations of $\alpha$ and $m$.
Values of $\gamma$ are evaluated by the least-squares fit.}
\end{figure}
The above analytical results are numerically verified in this
section. At first, we confirm the behavior of the degree
distribution function $P(k)$ for geographical networks on the
Sierpinski node set (Sierpinski geographical networks). Nodes
are located on vertices of the Sierpinski gasket with the
fractal dimension $D=\log 3/\log 2\approx 1.585$ and connected
by edges according to the condition Eq.~(\ref{condition1}) with
Eqs.~(\ref{Fxy}) and (\ref{sx}). In our numerical calculations
in this work, the lower cut-off $x_{\text{min}}$ is set to be
unity. A typical network on the 6th generation Sierpinski
gasket ($N=366$) is depicted in Fig.~1. There exist nodes
(hubs) possessing a large number of edges. Numerically
calculated degree distribution functions $P(k)$ for two
networks with different $m$ and $\alpha$ are presented in
Fig.~2. The scale-free property of networks formed by our
algorithm is clearly shown in this figure. The scale-free
exponents $\gamma$ calculated numerically for many Sierpinski
geographical networks with different combinations of $\alpha$
and $m$ are plotted in Fig.~3 as a function of $m(\alpha-1)/D$.
In this figure, values of $\gamma$ for networks with nodes
distributed homogeneously in two-dimensional Euclidean space
(2d-geographical networks) are also plotted. We should remark
that our theoretical arguments are valid for a geographical
network with homogeneously distributed nodes which is a special
case with the Euclidean dimension $d$ instead of the fractal
dimension $D$. In fact, Eq.~(\ref{gamma}) with $D=d$ reproduces
the results by \cite{morita2006,masuda2005}. We see that all
numerical results collapse onto the theoretical line given by
Eq.~(\ref{gamma}). These results strongly support the
theoretical predictions on $P(k)$ presented in
Sec.~\ref{Degree}.

\begin{figure}[ttt]
\includegraphics[width=0.4\textwidth]{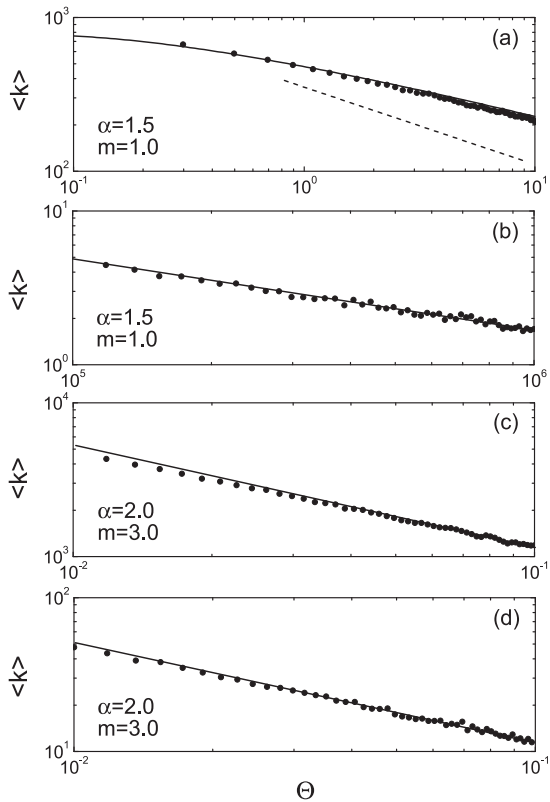}
\caption{\label{fig4}
Relation between the average degree $\langle k\rangle$ and the
threshold value $\Theta$ for (a) $m<m_{\text{c}0}$ and
$\Theta<\Theta_{0}$, (b) $m<m_{\text{c}0}$ and
$\Theta>\Theta_{0}$, (c) $m>m_{\text{c}0}$ and
$\Theta<\Theta_{0}$, and (d) $m>m_{\text{c}0}$ and
$\Theta>\Theta_{0}$. Dots represent results calculated
numerically for 2d-geographical networks. Details of parameters
and conditions are given in the text. Solid curves show the
theoretical predictions given by Eqs.~(\ref{kavresult1}) and
(\ref{kavresult2}). Dashed line in Fig.~4(a) indicates the
slope given by Eq.~(\ref{kavresult11}) without the logarithmic
term.}
\end{figure}
\begin{figure}[ttt]
\includegraphics[width=0.5\textwidth]{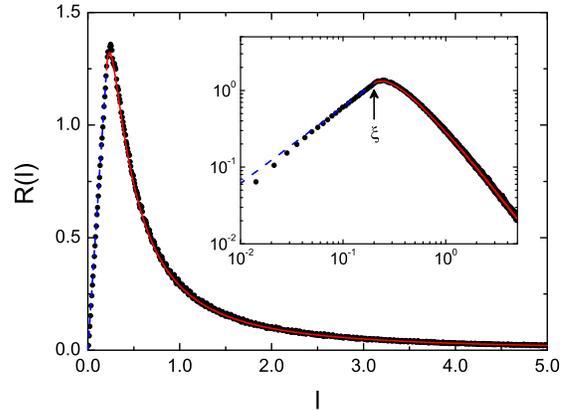}
\caption{\label{fig5}
(Color online) Edge-length distribution function $R(l)$ for
2d-geographical networks in squares of size $L=10.0$. In order
to obtain the numerical results (dots), we employ $\alpha=2.0$
($m_{\text{c}0}=2.0$), $m=3.0$, and $N=1,000$. The threshold
value $\Theta$ is chosen as $\Theta=125.2$ so that $\langle
k\rangle$ becomes equal to $10.0$. The results are averaged
over $1,000$ realizations. Solid and dashed lines represent the
theoretical results given by Eq.~(\ref{rl2}) and (\ref{rl3})
with the common prefactor presented by Eq.~(\ref{C}). The inset
shows the same result in a logarithmic scale. The length
$\xi(=0.20)$ defined by Eq.~(\ref{xi}) is indicated by the
arrow.
}
\end{figure}
Next, we confirm numerically the relation between $\langle
k\rangle$ and $\Theta$ for 2d-geographical networks. We have to
evaluate $\Theta_{0}$ defined by Eq.~(\ref{theta0}) to
distinguish four regions with respect to $m$ and $\Theta$
corresponding to Eqs.~(\ref{kavresult1}) and
(\ref{kavresult2}). In addition, $\Theta$ must satisfy the
condition $\Theta_{\text{min}}\ll \Theta \ll
\Theta_{\text{max}}$, where $\Theta_{\text{min}}$ and
$\Theta_{\text{max}}$ are given by Eqs.~(\ref{thetamin}) and
(\ref{thetamax2}), respectively. Rewriting Eqs.~(\ref{theta0})
and (\ref{thetamax2}) by
$\Theta_{0}=N^{1/(\alpha-1)}x_{\text{min}}^{2}L^{-m}$ and
$\Theta_{\text{max}}=N^{[2D+m(\alpha-1)]/[D(\alpha-1)]}x_{\text{min}}^{2}L^{-m}$,
we can estimate $\Theta_{0}$ and $\Theta_{\text{max}}$ without
treating the boundary-condition dependent geometrical factor
$\rho\Omega/D$. Figures 4(a) and 4(b) show numerically
calculated $\langle k\rangle$ as a function of $\Theta$ for
$m<m_{\text{c}0}$. In these calculations, the exponent $m$
characterizing the strength of the geographical effect are
chosen to be $1.0$, and $N=1,000$ nodes are distributed in a
square of size $L=100.0$. The fitness $x$ is allocated to each
node according to the distribution function $s(x)$ given by
Eq.~(\ref{sx}) with $\alpha=1.5$. The square system has the
periodic boundary conditions in the $x$ and $y$ directions.
Since $m_{\text{c}0}=4.0$, $\Theta_{0}=10^{4}$,
$\Theta_{\text{min}}=0.01$, and $\Theta_{\text{max}}=3.2\times
10^{11}$ from these parameters, the conditions
$\Theta_{\text{min}}\ll \Theta<\Theta_{0}\ll
\Theta_{\text{max}}$ and $\Theta_{\text{min}}
\ll\Theta_{0}<\Theta\ll\Theta_{\text{max}}$ are satisfied for
Figs.~4(a) and 4(b), respectively, and $m<m_{\text{c}0}$ for
both figures. Numerically calculated $\langle k\rangle$s are
well described by solid lines representing
Eqs.~(\ref{kavresult11}) and (\ref{kavresult21}), where the
constant $c$ in Eq.~(\ref{kavresult11}) and prefactors are
suitably chosen. Results for $m>m_{\text{c}0}$ are presented in
Figs.~4(c) and 4(d) which show the $\langle k\rangle$-$\Theta$
relation for $\Theta<\Theta_{0}$ and $\Theta>\Theta_{0}$,
respectively. As in the cases of Figs.~4(a) and 4(b), nodes are
distributed in a square of size $L=100.0$. Parameters $\alpha$
and $m$ are chosen as $\alpha=2.0$ and $m=3.0$ so that the
condition $m>m_{\text{c}0}(=2.0)$ is satisfied. In order to
realize the condition $\Theta_{\text{min}}\ll \Theta<
\Theta_{0}\ll\Theta_{\text{max}}$ in Fig.~4(c), we treated a
large network with $N=100,000$, which has
$\Theta_{\text{min}}=10^{-6}$, $\Theta_{\text{max}}=3.2\times
10^{11}$, and $\Theta_{0}=0.1$. On the contrary, the number of
nodes for Fig.~4(d) is $N=1,000$, in which
$\Theta_{\text{min}}\ll
\Theta_{0}<\Theta\ll\Theta_{\text{max}}$ is satisfied with
$\Theta_{\text{min}}=10^{-6}$, $\Theta_{\text{max}}=3.2\times
10^{4}$, and $\Theta_{0}=0.001$. Similarly to Figs.~4(a) and
4(b), numerical results agree well with the theoretical results
shown by solid lines.

Finally, the behavior of the edge-length distribution function
$R(l)$ is examined also for 2d-geographical networks. Results
(dots) shown in Fig.~5 are calculated in the condition of
$m>m_{\text{c}0}$, for which $R(l)$ can be properly normalized
even in the thermodynamic limit. Solid and dashed curves
represent the theoretical prediction given by Eq.~(\ref{rl2})
and (\ref{rl3}) with the common prefactor presented by
Eq.~(\ref{C}). It should be emphasized that there is no fitting
parameter to obtain the theoretical curve. Numerical results
agree quite well with the theoretical prediction. The peak
structure and the tail profile slightly deviating from a power
law (see the inset of Fig.~5) result from the logarithmic term
of Eq.~(\ref{rl2}). The reason of the slight deviation between
numerical results and the theoretical line for $l\lesssim 0.05$
(see the inset) is due to a finite $\Delta l$ defined below
Eq.~(\ref{thetamax1}).

\section{Compactness and efficiency}
\label{Compactness}

From the functional forms of $P(k)$ and $R(l)$, we can
immediately find the convergence property of the average degree
$\langle k\rangle$ and the average edge length $\langle
l\rangle$ in the thermodynamic limit. As argued at the end of
Sec.~\ref{Relation}, the quantity $\langle k\rangle$ for $m\le
m_{\text{c}0}$ diverges as $L^{D-m(\alpha-1)}\log L$ and it
converges for $m>m_{\text{c}0}$. On the other hand, the
convergence of $\langle l\rangle$ is governed by
Eq.~(\ref{rl2}) describing $R(l)$ in the asymptotic $l$ regime.
From Eq.~(\ref{rl2}), the average $\langle l\rangle$ diverges
for $m\le m_{\text{c}1}$ and remains finite for $m>
m_{\text{c}1}$, where
\begin{equation}
m_{\text{c}1}=\frac{D+1}{\alpha-1}\ .
\end{equation}
The behavior of $\langle k\rangle$ and $\langle l\rangle$
suggests that the impact of the geographical effect can be
classified into three regions. For $0\le m\le m_{\text{c}0}$,
both quantities $\langle k\rangle$ and $\langle l\rangle$
diverge in the thermodynamic limit. Since a node connects with
a huge number of nodes far away we call this region the
\textit{non-compact phase}. In the same sense, the region of
$m>m_{\text{c}1}$ is termed the \textit{compact phase} where
both quantities $\langle k\rangle$ and $\langle l\rangle$
remain finite and a node connects with only a small number of
nodes in its vicinity. In the \textit{intermediate phase},
i.e., $m_{\text{c}0}<m\le m_{\text{c}1}$, the average degree
converges but $\langle l\rangle$ diverges. These regions are
summarize in Table~\ref{tab1}.
\begin{table}[ttt]
\caption{Three phases of geographical networks embedded in a fractal space.}
\begin{ruledtabular}
\begin{tabular}{llll}
Phase        & Region of $m$ & $\langle k\rangle$ & $\langle l\rangle$ \\
\hline
Non-compact  & $m\le m_{\text{c}0}$               & infinite & infinite \\
Intermediate & $m_{\text{c}0}<m\le m_{\text{c}1}$ & finite   & infinite \\
Compact      & $m> m_{\text{c}1}$                 & finite   & finite \\
\end{tabular}
\end{ruledtabular}
\label{tab1}
\end{table}

Let us consider the relation between the compactness of a network and
the geographical efficiency $e$ defined by
\begin{equation}
e=\frac{2}{N(N-1)}\sum_{i>j}\frac{1}{r_{ij}}\ ,
\label{e}
\end{equation}
where $r_{ij}$ is the shortest Euclidean distance between two
nodes $i$ and $j$ along a network path. This quantity is a
natural extension of the global efficiency $E$ defined by
\begin{equation}
E=\frac{2}{N(N-1)}\sum_{i>j}\frac{1}{d_{ij}}\ ,
\label{E}
\end{equation}
where $d_{ij}$ is the shortest network distance between two
nodes $i$ and $j$ introduced in \cite{latora2001}. Both
quantities $e$ and $E$ characterize the efficiency of the
information exchange or the flow in the network. If the
efficiency is governed mainly by the Euclidean distance along
the path rather than the number of steps, the geographical
efficiency $e$ is more meaningful than $E$, and vice versa.

\begin{figure}[ttt]
\includegraphics[width=0.55\textwidth]{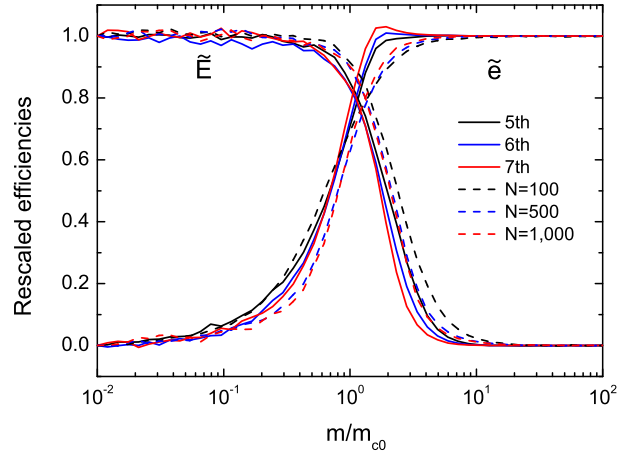}
\caption{\label{fig6}
(Color online) Rescaled geographical and global efficiencies as
a function of $m/m_{\text{c}0}$ for Sierpinski (solid lines) and
2d- (dashed lines) geographical networks. For both network systems,
we employ $\alpha=2.0$ and $\Theta$ giving $\langle k\rangle=10.0$,
and all results are averaged over $1,000$ realizations. Numbers of
nodes in Sierpinski geographical networks in the 5th, 6th, and
7th generations are $123$, $336$, and $1,095$, respectively.
}
\end{figure}
We calculated numerically these quantities and examined how the
compactness is related to the efficiencies $e$ and $E$. Solid
lines in Fig.~6 represent the efficiencies for Sierpinski
geographical networks in three different generations (having
the same linear dimension $L=1.0$), while dashed lines show them
for 2d-geographical networks with different numbers of nodes in
squares of size $L=10.0$. In order to eliminate the $L$ and $N$
dependence of the maximum and minimum values of the
efficiencies, we plot the rescaled quantities $\tilde{e}$ and
$\tilde{E}$ defined by $\tilde{e}=(e-e_{\text{min}})/
(e_{\text{max}}-e_{\text{min}})$ and
$\tilde{E}=(E-E_{\text{min}})/(E_{\text{max}}-E_{\text{min}})$,
where $e_{\text{min}}$ ($E_{\text{min}}$) and
$e_{\text{max}}$ ($E_{\text{max}}$) are $e(E)$ at $m/m_{\text{c}0}=10^{-2}$ and
$10^{2}$, respectively. The rescaled geographical (global)
efficiency $\tilde{e}$ ($\tilde{E}$) rapidly increases
(decreases) near $m=m_{\text{c}0}$ as increasing $m$. From the
fact that enlarging the system the slope near $m=m_{\text{c}0}$
becomes steeper and the point at the intersection of
$\tilde{e}$ and $\tilde{E}$ approaches $m=m_{\text{c}0}$, the
efficiencies $e$ and $E$ in the thermodynamic limit are
supposed to show step-like forms at $m=m_{\text{c}0}$. This
behavior can be interpreted as follows. In the non-compact
phase ($m<m_{\text{c}0}$), existing short-cut edge (in the
topological sense) connects a starting node and a target node
by a small number of edges with going back-and-forth around the
target node. This back-and-forth motion, however, requires an
extra Euclidean distance and leads the low geographical
efficiency. On the other hand, in the compact phase, the
network structure resembles the structure of a regular lattice
(or regular Sierpinski gasket). Although we need lots of edges
to connect two distant nodes, the total Euclidean length along
the path can be minimized as the geodetic distance between two
nodes. Thus, the geographical efficiency $e$ becomes large in
this region. The above consideration supports that the global
efficiency $E$ measuring the number of edges to connect nodes
is large for $m<m_{\text{c}0}$ and small for $m>m_{\text{c}0}$.
It should be noted that the abrupt change in $e$ or $E$ occurs
at $m=m_{\text{c}0}$ but not at $m=m_{\text{c}1}$ though the
transition point $m_{\text{c}1}$ is related to the edge length.
Since $e$ and $E$ behave oppositely, it is crucial to clarify
which efficiency is more relevant to a given problem by
considering how strongly the cost of the flow is influenced by
the Euclidean distance. It is also interesting that both
efficiencies $e$ and $E$ are relatively high in the network at
$m=m_{\text{c}0}$ at which the competition between order and
disorder in the geometrical sense is balanced.

\section{Example}
\label{Example}

\begin{figure}[ttt]
\includegraphics[width=0.35\textwidth]{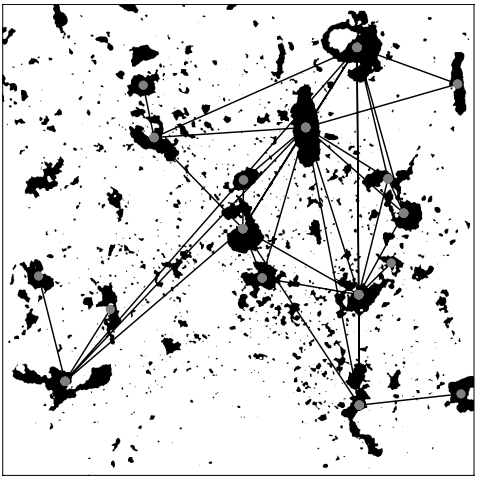}
\caption{\label{fig7}
2d X-ray CT image of a soil structure with network edges
connecting pores (black patches) according to
Eq.~(\ref{condition1}) with $m=5$ and $\Theta=0.01$. From
\cite{mooney2009}.
}
\end{figure}
\begin{figure}[ttt]
\includegraphics[width=0.38\textwidth]{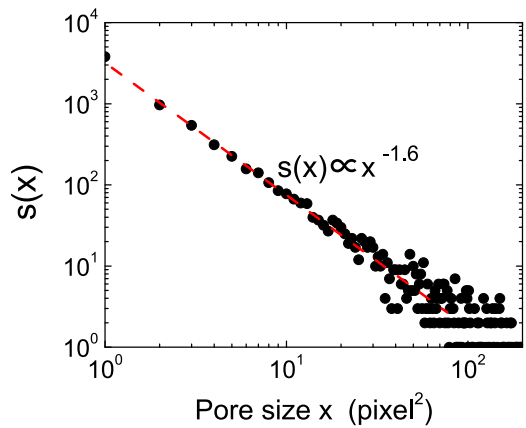}
\caption{\label{fig8}
(Color online) Pore-size distribution obtained from the image
analysis of the soil structure shown in Fig.~7. Dashed line
shows the power-law fit $s(x)\sim x^{-\alpha}$ to the
distribution with $\alpha=1.6$.
}
\end{figure}
\begin{figure}[bbb]
\includegraphics[width=0.48\textwidth]{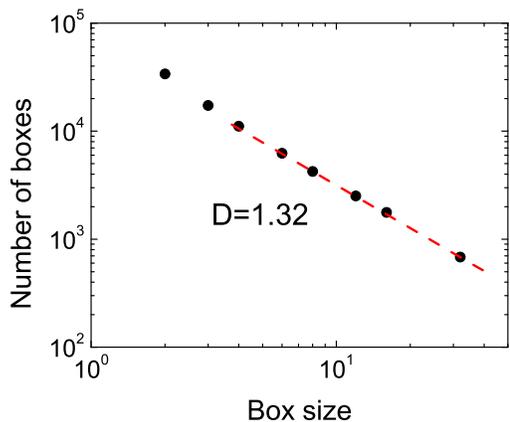}
\caption{\label{fig9}
(Color online) Number of boxes required to cover the soil-pore
image shown in Fig.~7 as a function of the box size. The slope
obtained by the least-squares fit (dashed line) indicates that
the fractal dimension of the soil-pore structure is $1.32$.
}
\end{figure}
We will demonstrate the above described approach using soil-pore
networks as an example of a real-world spatially embedded
network. Two approaches to build complex network models of
soil-pore organization have recently been developed
\cite{mooney2009,santiago2008,cardenas2010}. Here, we will
concentrate on the networks presented in \cite{mooney2009}
formed by geographical threshold algorithm as described in
Sec.~\ref{Model}. We consider a set of $N$ pores representing
the nodes of the network. The nodes of the network are the
centers of the pores and the links between nodes are drawn
according to Eq.~(\ref{condition1}) with Eqs.~(\ref{Fxy}) and
(\ref{sx}), where in this case the continuous fitness variable
$x$ is the size of the pore. Pore sizes and their relative
positions of actual soil specimens are obtained from the image
analysis of 2d soil X-CT scans \cite{mooney2009}.

An example of the soil-pore network overlain on the 2d soil
porous structure image is presented in Fig.~7 (from
\cite{mooney2009}). The pore-size distribution of the soil
sample shown in Fig.~7 is analyzed to be of the form $s(x)\sim
x^{-\alpha}$ with $\alpha=1.6$ (Fig.~8). The fractal dimension
$D=1.32$ of the soil-pore structure is also determined using
box-counting method as shown in Fig.~9. Our theory predicts
that the soil-pore network has a power-law degree distribution
function and the degree exponent $\gamma$ depends on $m$ above
$m_{\text{c}0}=2.2$. Figure 10 shows the degree distributions
for three networks based on soil image data constructed with
different values of the parameter $m$ ($m=1$, $3$, and $5$). In
this figure, the power-law fits (dashed lines) and calculated
exponents of the degree distributions are shown together with
theoretically predicted scale-free exponents from
Eq.~(\ref{gamma}). The results clearly demonstrate the
agreement of the empirically obtained scale-free exponents
and the theoretically predicted ones. This also
shows how the fractality of pore spatial distribution is
reflected in the network organization. By measuring the scaling
exponent $\gamma$ of the degree distribution for any $m>m_{c0}$
and the exponent $\alpha$ of the pore-size distribution, the
fractal dimension $D$ can be determined.
\begin{figure}[ttt]
\includegraphics[width=0.42\textwidth]{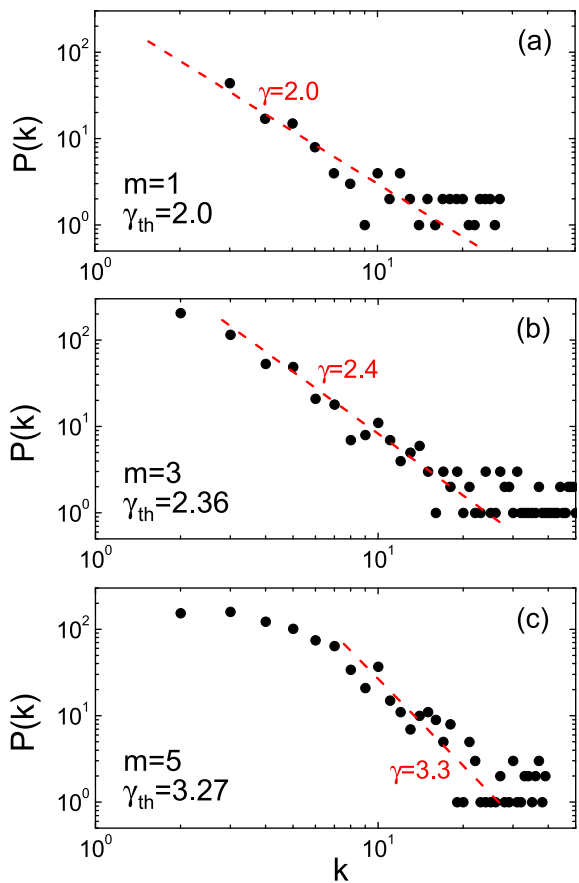}
\caption{\label{fig10}
(Color online) Degree distributions for three soil-pore networks
formed by (a) $m=1$, (b)$m=3$, and (c) $m=5$. Dashed lines through
dots represent the power-law fits with the slopes indicated beside
the lines. $\gamma_{\text{th}}$ in each panel gives the degree
exponent predicted by Eq.~(\ref{gamma}) with $D=1.32$ and $\alpha=1.6$.
}
\end{figure}

\section{Conclusions}
\label{Conclusions}

We have proposed a geographical scale-free network model with
the nodes embedded in a fractal space and analytically and
numerically studied several network properties. The fractal
dimension $D$ of the embedding space was found to influence the
scale-free exponent as $\gamma=m(\alpha-1)/D+1$ only if the
spatial embedding is strong enough (i.e., when $m>m_{c0}$)
otherwise $\gamma=2$. The analyses of the average degree and
average edge length revealed that this type of network can
exist either in the non-compact, compact, or intermediate phase
depending on the importance of the spatial arrangement of
nodes. We derived the edge-length distribution functions for
our network model and showed that it has a peak-like structure
similar to the profile of the shortest-path-distance
distribution observed in a large-scale structure of the
Internet \cite{vazquez2002,mahadevan2006}. It is interesting to
apply our approach to modeling the Internet at the autonomous
systems level considering the observed long-tailed distribution
of autonomous systems sizes \cite{lakhina2003}. The measured
degree distribution exponent of the Internet is slightly larger
that $\gamma=2$ ($\gamma=2.1\sim 2.2$)
\cite{vazquez2002,siganos2003} and seems to be decreasing with
time \cite{siganos2003}. In our network model this would be an
indication of the evolution of the Internet towards the
non-compact phase ($m\to m_{c0}$).

We hope that our work will help in advancing the understanding
of the complex systems in which the heterogeneity of intrinsic
properties and the spatial arrangement of the elements play an
important role.

\begin{acknowledgments}
This work was supported in part by a Grant-in-Aid for
Scientific Research (No.~22560058) from Japan Society for the
Promotion of Science and by grant No.~J3-2290 from Slovenian Research Agency. 
Numerical calculations in this work were
performed on the facilities of the Supercomputer Center,
Institute for Solid State Physics, University of Tokyo.
\end{acknowledgments}

\end{document}